\documentclass[12pt,a4paper]{article}
\usepackage[utf8]{inputenc}
\usepackage{epsfig}
\usepackage{mathtools}
\usepackage{amsfonts}
\usepackage{amssymb}
\usepackage{amsmath}
\usepackage{cancel}
\usepackage{color}
\usepackage{slashed}
\usepackage{cite}
\usepackage{caption}
\usepackage{subcaption}
\usepackage{comment}
\usepackage{marginnote}

\usepackage{tikz}
\usetikzlibrary{shapes}
\usetikzlibrary{trees}
\usetikzlibrary{calc}
\usetikzlibrary{decorations.pathmorphing}
\usetikzlibrary{decorations.markings}

\def\FF              {\mathcal{F}}

\def\bP{{\bf P}}

\def\sign{\text{sgn}}

\def\s{\sigma}

\def\bp{{\bf p}}
\def\btp{{\bf\tilde p}}
\def\btP{{\bf\tilde P}}
\def\bP{{\bf P}}

\def\Z{\mathcal{Z}}
\def\P{\mathcal{P}}

\newcommand{\barH}{{\bar H}}

\begin{document}

\begin{center}

\vspace{1cm}

{ \Large\bf On analytical perturbative solution of ABJM quantum spectral curve} \vspace{1cm}

{\large R.N. Lee$^{1}$ and  A.I. Onishchenko$^{2,3,4}$}\vspace{0.5cm}

{\it $^1$ Budker Institute of Nuclear Physics, Novosibirsk,
	Russia,\\
	$^2$Bogoliubov Laboratory of Theoretical Physics, Joint
	Institute for Nuclear Research, Dubna, Russia, \\
	$^3$Moscow Institute of Physics and Technology (State University), Dolgoprudny, Russia\\
	$^4$Skobeltsyn Institute of Nuclear Physics, Moscow State University, Moscow, Russia}\vspace{1cm}

\abstract{Recently we showed how non-homogeneous second-order difference equations appearing within ABJM quantum spectral curve description could be solved using Mellin space technique. In particular we provided explicit results for anomalous dimensions of twist 1 operators in $sl(2)$ sector at arbitrary spin values up to four loop order. It was shown that the obtained results may be expressed in terms of harmonic sums decorated by fourth root of unity factors, so that maximum transcendentality principle holds. In this note we show that the same result could be also obtained by direct solution of the mentioned equations in spectral parameter $u$-space. The solution involves new highly nontrivial identities between hypergeometric functions, which may have various other applications. We expect this method to be more easily generalizable to higher loop orders as well as to other theories, such as $\mathcal{N}=4$ SYM.}
\end{center}

\begin{center}
	ABJM, Baxter equations	
\end{center}

\newpage

\tableofcontents{}\vspace{0.5cm}

\renewcommand{\theequation}{\thesection.\arabic{equation}}

\section{Introduction}

Recently following the discovery of AdS/CFT duality \cite{tHooftDuality,MaldacenaAdSCFT,GKP,WittenAdSHolography}
we have seen a lot of progress in understanding the integrable structures behind the quantum field theories with extended supersymmetry in dimensions greater then two, see for a review and introduction \cite{IntegrabilityReview,IntegrabilityPrimer,IntegrabilityDeformations,IntegrabilityDefects,IntroductionQSC,IntegrabilityStructureConstants,FishnetCFTreview}. The most well understood theories are given by $\mathcal{N}=4$ SYM in four and $\mathcal{N}=6$ super Chern-Simons theory (ABJM model) in three dimensions \cite{ABJM1}. It was shown, that different techniques from the world of integrable systems, such as worlsheet and spin-chain S-matrices \cite{StaudacherSMatrix,BetheAnsatzQuantumStrings,BeisertDynamicSmatrix,BeisertAnalyticBetheAnsatz,TranscedentalityCrossing,JanikWorldsheetSmatrix,ArutyunovFrolovSmatrix,ZamolodchikovFaddevAlgebraAdS,N6Smatrix}, Asymptotic Bethe Ansatz (ABA) \cite{MinahanZarembo,N4SuperSpinChain,LongRangeBetheAnsatz,TranscedentalityCrossing,MinahanZaremboChernSimons,SpinChainsN6ChernSimons,AllloopAdS4} and Thermodynamic Bethe Ansatz (TBA) \cite{TBAN4,TBAN4proposal,TBAexcitedstates,TBAMirrorModel} as well as $Y$ and $T$-systems \cite{YsystemAdS5,TBAfromYsystem,WronskianSolution,SolvingYsystem,TBAYsystemAdS4,GromovYsystemAdS4,DiscontinuityRelationsAdS4} are applicable for the computation of conformal spectrum of these theories. The integrability based methods were also used in the study of quark-antiquark potential \cite{potentialTBA,IntegrableWilsonLoops,cuspQSC,potentialQSC}, expectation values of polygonal Wilson loops at strong coupling and beyond \cite{BubbleAnsatz,YsystemScatteringAmplitudes,OPEpolygonalWilsonLines,SmatrixFiniteCoupling,OPEHelicityAmplitudes,AsymptoticBetheAnsatzGKPvacuum}, eigenvalues of BFKL kernel \cite{adjointBFKL,GromovBFKL1,GromovBFKL2,BFKLnonzeroConformalSpin}, structure constants \cite{StructureConstantsPentagons,StructureConstantsWrappingOrder,ClusteringThreePointFunctions,StructureConstantsLightRayOperators,QSC_StructureConstants}, correlation functions \cite{StructureConstantsPentagons,CorrelationFunctionsIntegrability1,CorrelationFunctionsIntegrability2,CorrelationFunctionsIntegrability3,CorrelationFunctionsIntegrability4,CorrelationFunctionsIntegrability5,CorrelationFunctionsIntegrability6}, one-point functions of operators in the defect conformal field theory \cite{defectCFT1,defectCFT2,defectCFT3} and thermal observables such as Hagedorn temperature of $\mathcal{N}=4$ SYM \cite{HagedornTemperatureIntegrability1,HagedornTemperatureIntegrability2}.

Further detailed study of TBA equations for  $\mathcal{N}=4$ SYM and ABJM models has led to the discovery of Quantum Spectral Curve (QSC) formulation for these models \cite{N4SYMQSC1,N4SYMQSC2,twistedN4SYMQSC,N4SYMQSC3,ABJMQSC,ABJMQSCdetailed,QSCetadeformed}. QSC is an alternative reformulation of TBA equations in terms of nonlinear Riemann-Hilbert problem. The numerical solution of QSC for these theories at finite coupling applicable even for complexified spin was presented in \cite{QSCnumericsN4SYM1,QSCnumericsN4SYM2,QSCnumericsABJM}. The iterative procedure for perturbative solution of the mentioned Riemann-Hilbert problems for these theories at weak coupling up to in principle arbitrary loop order was given in \cite{VolinPerturbativeSolution,ABJMQSC12loops}. The techniques presented there are however limited to the situation when the states quantum numbers are given explicitly as some integer numbers. They are sufficient for the recovery of full analytical structure of the conserved charges in the respective spin chains provided that we know a finite basis of functions in terms of which they could be written explicitly. When it is not the case we may use for example Mellin space technique to solve QSC equations without assigning specific integer values to state quantum numbers \cite{ABJM_QSC_Mellin}. In this note we want to show, that there is actually a way to solve QSC equations directly in spectral parameter $u$-space without imposing any restriction on state quantum numbers. In fact we will restrict ourselves here only to the case of twist 1 operators in $sl(2)$ sector of ABJM model up to four loop order and defer the generalization to higher loops and other states to forthcoming publications. The presented approach, which is still under development, has also the potential for generalization to the case of twisted $\mathcal{N}=4$ SYM and ABJM quantum spectral curves with different non-polynomial large spectral parameter asymptotics for functions entering corresponding Riemann-Hilbert problems. Moreover, similar ideas should be useful in the study of BFKL regime within QSC approach \cite{GromovBFKL1,GromovBFKL2,BFKLnonzeroConformalSpin}, which still employs perturbative expansion in coupling constant $g$ and parameter $w\equiv S+1$ describing proximity of Lorentz spin $S$ to $-1$, so that the ratio $g^2/w$ remains fixed.

The present paper is organized as follows.  In the next section we present the set of equations together with employed strategy used to find perturbative solution of ABJM quantum spectral curve in the case of twist 1 operators in $sl(2)$ sector.  Next, in section \ref{BaxterEquations} we present the details of solution of corresponding Baxter equations. Section \ref{AnomalousDimensions} contains the discussion of our results for anomalous dimensions of twist 1 operators up to four loop order. Finally, in section \ref{Conclusion} we come with our conclusion.

\section{Perturbative solution of ABJM QSC}\label{ABJM-QSC-PerturbativeSolution}

ABJM is a three-dimensional $\mathcal{N}=6$ Chern-Simons theory with product gauge group $U(N)\times \hat{U}(N)$ at levels $\pm k$. The field content of the theory consists of two gauge fields $A_{\mu}$ and $\hat{A}_{\mu}$, four complex scalars $Y^A$ and four Weyl spinors $\psi_A$. The matter fields transform in the bi-fundamental representation of the gauge group. The global symmetry group of ABJM theory for Chern-Simons level $k > 2$ is given by orthosymplectic supergroup $\text{OSp} (6|4)$ \cite{ABJM1,ABJM2} and the ``baryonic'' $U(1)_b$ \cite{ABJM2}. Here we will use as example anomalous dimensions of twist 1 $sl (2)$-like states given by single-trace operators of the form \cite{KloseABJMreview}:
\begin{eqnarray}
\text{tr} \left[ D_{+}^S (Y^1 Y_4^\dagger)\right] . \label{sl2op}
\end{eqnarray}
At present the quantum spectral curve (QSC) method is the most advanced method to deal with spin chain spectral problems arising in the study if $\text{AdS}_{d+1}/\text{CFT}_d$ duality. For the case of ABJM theory QSC formulation was introduced in \cite{ABJMQSC,ABJMQSCdetailed}, see also \cite{ABJMQSC12loops}. 

For the perturbative solution of ABJM quantum spectral curve we will use the same set of equations as in \cite{ABJMQSC12loops}. The latter easily follow\footnote{See \cite{ABJMQSC12loops} for details.} from Riemann-Hilbert problem for $\bP \nu$-system and is given in our case by
\begin{align}
\frac{\nu_1^{[3]}}{\bP_1^{[1]}} - \frac{\nu_1^{[-1]}}{\bP_1^{[-1]}} - \s \left(
\frac{\bP_0^{[1]}}{\bP_1^{[1]}} - \frac{\bP_0^{[-1]}}{\bP_1^{[-1]}}
\right)\nu_1^{[1]} &= -\s \left(
\frac{\bP_2^{[1]}}{\bP_1^{[1]}} - \frac{\bP_2^{[-1]}}{\bP_1^{[-1]}}
\right)\nu_2^{[1]}\, , \label{Baxternu1} \\
\frac{\nu_2^{[3]}}{\bP_1^{[1]}} - \frac{\nu_2^{[-1]}}{\bP_1^{[-1]}} + \s \left(
\frac{\bP_0^{[1]}}{\bP_1^{[1]}} - \frac{\bP_0^{[-1]}}{\bP_1^{[-1]}}
\right)\nu_2^{[1]} &= \s \left(
\frac{\bP_3^{[1]}}{\bP_1^{[1]}} - \frac{\bP_3^{[-1]}}{\bP_1^{[-1]}}
\right)\nu_1^{[1]}\, , \label{Baxternu2}
\end{align}
and 
\begin{align}
\s \nu_1^{[2]} &= \bP_0 \nu_1 - \bP_2 \nu_2 + \bP_1 \nu_3 \, , \label{nu3sol} \\
\s \nu_2^{[2]} &= -\bP_0 \nu_2 + \bP_3 \nu_1 + \bP_1 \nu_4 \, , \label{nu4sol} \\
\btP_2 - \bP_2 &= \s \left(
\nu_3 \nu_1^{[2]} - \nu_1 \nu_3^{[2]}
\right)\, , \label{pt2} \\
\btP_1 - \bP_1 &= \s \left(
\nu_2 \nu_1^{[2]} - \nu_1 \nu_2^{[2]}
\right)\, , \label{pt1} \\
\left(
\nu_1 + \s \nu_1^{[2]}
\right) \left(
\bp_0 - h x
\right) &= \bp_2 \left(
\nu_2 + \s \nu_2^{[2]}
\right) - \bp_1 \left(
\nu_3 + \s \nu_3^{[2]}
\right)\, , \label{p0}
\\ 
\left(
\nu_2 + \s \nu_2^{[2]}
\right)\left(
\bp_0 + h x
\right) &= \bp_3 \left(
\nu_1 + \s \nu_1^{[2]}
\right) + \bp_1 \left(
\nu_4 + \s \nu_4^{[2]}
\right)\, ,  \label{p3} \\
( \bP_0 )^2 &= 1 - \bP_1 \bP_4 + \bP_2 \bP_3 . \label{P0}
\end{align}
where $\bP_i = (x h)^{-1} \bp_i$ and
\begin{equation}
x\equiv x (u) = \frac{u+\sqrt{u^2 - 4 h^2}}{2 h}
\end{equation}
is the Zhukovsky variable used to parametrize single cut of $\bP$ functions on the defining Riemann sheet. In addition from the analytical structure of $\nu_i (u)$ functions on the defining Riemann sheet it follows, that the following combinations of functions
\begin{align}
\nu_i (u) + \tilde{\nu}_i (u) &= \nu_i (u) + \s \nu_i^{[2]} (u)\, , \nonumber \\
\frac{\nu_i (u) - \tilde{\nu}_i (u)}{\sqrt{u^2 - 4 h^2}} &=
\frac{\nu_i (u) - \s \nu_i^{[2]} (u)}{\sqrt{u^2 - 4 h^2}}\, 
\label{cutsfree}
\end{align}
are free of cuts on the whole real axis. Similar to \cite{VolinPerturbativeSolution,ABJMQSC12loops} we will look for the solution employing the following ansatz of $\bP (u)$ functions\footnote{ABJM QSC coupling constant $h$ is a nontrivial function of ABJM t'Hooft coupling constant $\lambda$ \cite{SpinChainsN6ChernSimons,StringDualN6ChernSimons}, which scales as $h\sim\lambda$ at small and as $h\sim\sqrt{\lambda /2}$ at strong coupling constant.}:
\begin{align}
\bP_1 &= (x h)^{-1} \bp_1 = (x h)^{-1} \left(
1+\sum_{k=1}^{\infty}\sum_{l=0}^{\infty} c_{1,k}^{(l)}\frac{h^{2 l+k}}{x^k}
\right)\, , \\
\bP_2 &= (x h)^{-1} \bp_2 = (x h)^{-1}\left(
\frac{h}{x} + \sum_{k=2}^{\infty}\sum_{l=0}^{\infty} c_{2,k}^{(l)} \frac{h^{2 l+k}}{x^k}
\right)\, , \\
\bP_0 &= (x h)^{-1} \bp_0  = (x h)^{-1} \left(
\sum_{l=0}^{\infty} A_0^{(l)} h^{2 l} u + 
\sum_{l=0}^{\infty} m_0^{(l)} h^{2 l}
+\sum_{k=1}^{\infty}\sum_{l=0}^{\infty} c_{0,k}^{(l)}
\frac{h^{2 l+k}}{x^k}
\right)\, , \\
\bP_3 &= (x h)^{-1} \bp_3 = (x h)^{-1}\left(
\sum_{l=0}^{\infty} A_3^{(l)} h^{2 l} u^{3}
+ \sum_{j=0}^{2}\sum_{l=0}^{\infty} k_j^{(l)} h^{2 l} u^j
+ \sum_{k=1}^{\infty}\sum_{l=0}^{\infty} c_{3,k}^{(l)}\frac{h^{2 l+k}}{x^k}
\right) \, .
\end{align}
where we have accounted for correct large $u$ asymptotic of $\bP$ functions \cite{ABJMQSC,ABJMQSCdetailed,ABJMQSC12loops}, playing the role of initial conditions for the above equations depending on quantum numbers of operators we are interested in 
\begin{eqnarray}
\bP_a &\simeq& ( A_1 u^{-1}, \,A_2 u^{-2}, \,A_3 u^{2}, \,A_4 u, \, A_0 u^0) , \nonumber\\
A_1 A_4 &=&-\frac{(\Delta -1+S) (\Delta -S) (\Delta +2-S) (\Delta +1+S)}{3} \nonumber\\
A_2 A_3 &=&-\frac{(\Delta +S-2) (\Delta -1-S) (\Delta -S+3) (\Delta +S+2)}{12},\label{Pasympt}
\end{eqnarray}
The anomalous dimension $\gamma$, which we are interested in, is given by $\gamma = \triangle - 1 - S$. The coefficients $A_0^{(l)}$, $A_3^{(l)}$, $c_{i,k}^{(l)}$,  $m_j^{(l)}$ and $k_j^{(l)}$ are some functions of spin $S$ only, otherwise they are just constants. Here we have also used the  gauge freedom\footnote{See \cite{ABJMQSC12loops} for details.} to set $A_1 = 1$ and $A_2 = h^2$. The analytically continued though the cut functions are defined as
\begin{equation}
\btP_a = \left(\frac{x}{h}\right)^L \btp_a\, , \quad \btp_a = \bp_a\Big|_{x\to 1/x}\, .
\end{equation}
and the expansion of $\nu_i (u)$ functions in terms of QSC coupling constant $h$ is given by
\begin{equation}
\nu_{i} (u) = \sum_{l=0}^{\infty} h^{2 l - L} \nu_i^{(l)} (u)
\end{equation}

\section{Solution of  Baxter equations}\label{BaxterEquations}

The most complicated part of the QSC solution is the solution of two inhomogeneous Baxter equations at each perturbation order. To solve these second order finite difference equations at arbitrary values of spin $S$  values in \cite{ABJM_QSC_Mellin} we employed Mellin transform technique to convert difference equations to ordinary differential equations. Here, we want to show that there is actually a simpler way to solve mentioned Baxter equations directly in spectral parameter $u$-space. In order to iteratively search for the perturbative solution of equations \eqref{Baxternu1}--\eqref{p3}, we expand Eqs. \eqref{Baxternu1}, \eqref{Baxternu2} up to $h^k$ and obtain the inhomogeneous equations for $q_{1,2}^{(k)}=\left(\nu_{1,2}^{(k)}\right)^{[-1]}$ in the following form
\begin{align}
(u+i/2)q_{1}^{(k)}(u+i)-i(2S+1)q_{1}^{(k)}(u)-(u-i/2)q_{1}^{(k)}(u-i)=&V_{1}^{(k)} \label{Baxter1a}\,,\\
(u+i/2)q_{2}^{(k)}(u+i)+ i(2S+1)q_{2}^{(k)}(u)-(u-i/2)q_{2}^{(k)}(u-i)=&V_{2}^{(k)}\,. \label{Baxter2a}
\end{align}
Here $V_{1}^{(k)}$ depends on $q_{1,2}^{(l)}$ with $l<k$, and $V_{2}^{(k)}$ depends in addition on on $q_1^{(k)}$. The solution of homogeneous parts of Baxter equations was described in \cite{ABJM_QSC_Mellin} and is given by
\begin{align}
q_1^{hom} (S,u) &= \Phi_{1}^{per} Q_S (u) + \Phi_{1}^{anti} \Z_S (u)\, , \\
q_2^{hom} (S,u) &= \Phi_{2}^{anti} Q_S (u) + \Phi_{2}^{per} \Z_S (u)\, ,
\end{align}
where ($\sigma = (-1)^S$)
\begin{align}
Q_S\left(u\right) =&\, \frac{(-1)^{S}\Gamma\left(\frac{1}{2}+iu\right)}{S!\Gamma\left(\frac{1}{2}+iu-S\right)}\,_{2}F_{1}\left(-S,\frac{1}{2}+iu;\frac{1}{2}+iu-S;-1\right)\, ,\label{eq:BaxterPolynomial} \\
\Z_S (u) =&\, i\sigma\sum_{k=0}^{\left\lfloor \frac{S-1}{2}\right\rfloor }\frac{1}{S-k}Q_{S-1-2k}\left(u\right)+\sigma\eta_{-1}(u+{i}/{2})Q_S\left(u\right)
\end{align}
and $\Phi_{i}^{per}$ and $\Phi_{i}^{anti}$  are arbitrary periodic and anti-periodic functions in spectral parameter $u$. Otherwise they are arbitrary functions of spin $S$ to be determined from consistency conditions mentioned in previous section. We will parametrize their $u$ dependence similar to \cite{VolinPerturbativeSolution,ABJMQSC12loops} with the basis of periodic and anti-periodic combinations of Hurwitz functions defined as 
\begin{equation}
\P_k (u) = \eta_k (u) + \sign(k)(-1)^k\eta_k (i-u) = \text{sgn}(k)\P_k (u+i)\, , \quad k\neq 0 \in \mathbb{Z}\, ,
\end{equation}
where 
\begin{equation}
\eta_a  (u) = \sum_{k=0}^{\infty}\frac{(\text{sgn}(a))^k}{(u + i k)^{|a|}}
\end{equation}
Then the functions  $\Phi_{a}^{\mathrm{per}}$ and $\Phi_{a}^{\mathrm{anti}}$ take the form
\begin{equation}
\Phi_{a}^{\mathrm{per}} (u) = \phi_{a,0}^{\mathrm{per}} + \sum_{j=1}^{\Lambda}\phi_{a,j}^{\mathrm{per}} \P_j (u)\, , \quad \Phi_{a}^{\mathrm{anti}} (u) =  \sum_{j=1}^{\Lambda}\phi_{a,j}^{\mathrm{anti}}\P_{-j} (u)
\end{equation}
where $\Lambda$ is a cutoff dependent on the order of perturbation theory.

To find a particular solution of inhomogeneous Baxter equation let's first rewrite original Baxter equations in a more convenient form. Substituting the ansatz $q_1^{(k)} (u) = Q_S(u) F_S^{(k)}(u+i/2)$ into the first Baxter equation  (\ref{Baxter1a}) we get
\begin{equation}
-\nabla_{-} \left(
u Q_S^{[1]} Q_S^{[-1]}\nabla_{+} F_S^{(k)}
\right) = Q_S^{[1]} V_1^{(k)[1]}
\end{equation}
where  $\nabla_{+} f = f - f^{[2]}$ and $\nabla_{-} f = f + f^{[2]}$. Introducing inverse operators $\Psi_{\pm}$, such that $\nabla_{\pm}\Psi_{\pm} f =  f$, and solving this difference equation for $F_S^{(k)}(u)$ we find
\begin{equation}
F^{(k)} = - \Psi_{+} \left(\frac{1}{u Q_S^{[1]} Q_S^{[-1]}}\Psi_{-}\left(Q_S^{[1]} V_1^{(k) [1]}\right)\right)\, .
\end{equation}
Now, using the remarkable relation\footnote{We would like to stress that initially we have actually guessed this identity. It was used in \cite{ABJM_QSC_Mellin} to find a second solution of homogeneous Baxter equation, where in Appendix B also a rigorous proof that the latter is actually the desired solution was presented. We would like to mention that similar identity could be written for twist $2$ operators.} (see its proof in the  Appendix \ref{elemop})
\begin{multline}\label{eq:Qid}
\frac{1}{u Q_S^{[1]} Q_S^{[-1]}} = \frac{(-1)^S}{u} + i (-1)^S \sum_{k=0}^{\left[\frac{S-1}{2}\right]}\frac{1}{S-k} \left(
\frac{Q_{S-1-2k}^{[-1]}}{Q_S^{[-1]}} + \frac{Q_{S-1-2k}^{[1]}}{Q_S^{[1]}} \right) \\
= (-1)^S \nabla_{-} \left(
\eta_{-1} (u) + i \sum_{k=0}^{\left[\frac{S-1}{2}\right]}\frac{1}{S-k} \frac{Q_{S-1-2k}^{[-1]}}{Q_S^{[-1]}}
\right)
\end{multline}
the expression for $F_S^{(k)}(u)$ takes the form 
\begin{equation}
F_S^{(k)}(u) = - (-1)^S \Psi_{+} \Bigg\{ 
\nabla_{-} \Bigg(
\eta_{-1} (u) + i \sum_{k=0}^{\left[\frac{S-1}{2}\right]}\frac{1}{S-k} \frac{Q_{S-1-2k}^{[-1]}}{Q_S^{[-1]}}
\Bigg)\Psi_{-} (Q_S^{[1]} V_1^{(k) [1]})
\Bigg\}\, .
\end{equation}
Next, using the relation\footnote{It could be easily derived using summation by parts.} 
\begin{equation}
\nabla_{-} f^{[-1]}\Psi_{-} g^{[1]} = -\nabla_{+} \left(
f\Psi_{-} g
\right)^{[-1]} + (f g)^{[-1]}
\end{equation}
we get 
\begin{multline}
F_S^{(k)}(u) = - (-1)^S \Psi_{+} \left(
\frac{1}{u}\Psi_{-} \left(Q_S V_1^{(k)}\right)^{[1]}\right) \\ + (-1)^S \frac{P_S^{[-1]}}{Q_S^{[-1]}}\Psi_{-}\left(Q_S V_1^{(k)}\right)^{[-1]} - (-1)^S \Psi_{+} \left(P_S V_1^{(k)}\right)^{[-1]}
\end{multline}
where the $P_S (u)$ function is given by 
\begin{equation}
P_S (u) = i  \sum_{k=0}^{\left[\frac{S-1}{2}\right]}\frac{1}{S-k} Q_{S-1-2k}(u)
\end{equation}
Finally, for $q_1^{(k)}(u)$  we get
\begin{multline}
q_1^{(k)}(u) = - (-1)^S Q_S \Psi_{+} \left(
\frac{1}{u+i/2}\Psi_{-} \left(Q_S V_1^{(k)}\right)^{[2]}\right) \\
+ (-1)^S P_S\Psi_{-}\left(Q_S V_1^{(k)}\right) - (-1)^S Q_S \Psi_{+} \left(P_S V_1^{(k)}\right)
\end{multline}
Introducing $\FF_1^S$ operator as 
\begin{equation}
\FF_1^S [f] = - Q_S \Psi_{+} \left(
\frac{1}{u+i/2}\Psi_{-} \left(Q_S (-1)^S f\right)^{[2]}\right) - Q_S \Psi_{+} \left(
P_S (-1)^S f
\right) + P_S \Psi_{-}\left(Q_S (-1)^S f\right)
\end{equation}
the general solution of the first Baxter equation takes the form 
\begin{equation}
q_1^{(k)} = \FF_1^S \left[V_1^{(k)}\right] + Q_S \Phi_1^{per, (k)} + \Z_S  \Phi_1^{anti, (k)} \, .
\end{equation}
Similarly for the second Baxter equation we get 
\begin{equation}
q_2^{(k)} = \FF_2^S \left[V_1^{(k)}\right] + Q_S \Phi_2^{anti, (k)} + \Z_S  \Phi_2^{per, (k)} \, ,
\end{equation}
where
\begin{equation}
\FF_2^S [f] = - Q_S \Psi_{-} \left(
\frac{1}{u+i/2}\Psi_{+} \left(Q_S (-1)^S f\right)^{[2]}\right) + Q_S \Psi_{-} \left(
P_S (-1)^S f
\right) - P_S \Psi_{+}\left(Q_S (-1)^S f\right)\, .
\end{equation}
It turns out that, at least, up to NLO\footnote{The inhomogeneity in this case is polynomial or polynomial times $\eta_1$. The generalization to arbitrary loop order will be considered in one of our future publications.} the right hand side of Baxter equations can be transformed to the form containing only Baxter polynomials $Q_S(u)$ for different spin values $S$ and products of Baxter polynomials with $\eta_1$ - function. For the actions of $\FF_{1,2}^S$ operators on Baxter polynomials we find
\begin{align}
\FF_1^{S_1} \left[Q_{S_2}\right] &=  -\frac{i}{2} \frac{Q_{S_1}-Q_{S_2}}{S_1 - S_2}\, , \quad S_1 \neq S_2\, \label{eq:FQaction}\\
\FF_1^S \left[
Q_S
\right] &= -\frac{1}{2} Q_S\,  \eta_1 (u+i/2) - \frac{i}{2}\sum_{k=1}^S \frac{1+(-1)^k}{k} Q_{S-k}\, , \\
\FF_2^{S_1} \left[Q_{S_2}\right] &=  -\frac{i}{2} \frac{Q_{S_2}}{S_1 + S_2 + 1}\, .
\end{align}
In addition we need one extra rule for $\FF_2^S$:
\begin{align}
\FF_2^{S_1} \left[\eta_{1}^{[1]} Q_{S_2}\right] &= \frac{1}{2 i (S_1 + S_2 +1)}\left\{
\eta_1^{[1]} Q_{S_2} + \FF_2^{S_1} \left[Q_{S_2}^{[2]}\right] + \FF_2^{S_1} \left[Q_{S_2}^{[-2]}\right]
\right\}
\end{align}
The elementary operations needed to transform inhomogeneous part to the required form are given by\footnote{The derivation of Eqs. \eqref{eq:FQaction}-\eqref{eq:uQmult} will be presented elsewhere. However, in the course of the proof of Eq. \eqref{eq:Qid} we prove some of these identities in Appendix \ref{elemop}.}
\begin{equation}
\frac{Q_S}{u\pm i/2} = \frac{(\mp 1)^S}{u\pm i/2} - 2 i \sum_{k=1}^S (\pm 1)^{k+1} Q_{S-k}\sum_{l=0}^{k-1} \frac{(-1)^l}{S-l}\, ,
\end{equation}
\begin{equation}
Q_S^{[\pm 2]} = Q_S + 2\sum_{k=1}^S (\pm 1)^k Q_{S-k}
\end{equation}
and 
\begin{equation}\label{eq:uQmult}
u\, Q_S = \frac{i}{2} (S+1) Q_{S+1} - \frac{i}{2} S Q_{S-1}\, .
\end{equation}
The results of solution of mentioned Baxter equations up to NLO could be found in \cite{ABJM_QSC_Mellin}. It is instructive to see types of functions entering solutions for $q_{1,2}$ at different orders of perturbation theory. For $q_1^{(0,1)}$, for example, we get\footnote{See definition of harmonic sums in the next section.} 
\begin{equation}
q_1^{(0)} = \alpha Q_S\, , 
\end{equation}
\begin{multline}
q_1^{(1)} = 4\alpha B_1 (S) Q_S \left\{\gamma_E + \log 2 - i\eta_1 (u+i/2) - H_1 (S) + i\pi\tanh (\pi u)\right\} \\+ \alpha \sum_{k=1}^S
\frac{1+(-1)^k}{k}\left(
3 B_1 (S) - B_1 (S-k)
\right)Q_{S-k} + \phi_{1,0}^{\mathrm{per}} Q_S
+ \phi_{1,1}^{\mathrm{per}} \P_1 (u+i/2) Q_S\, ,
\end{multline}
where
\begin{equation}
\alpha^2 = \frac{i}{4 B_1 (S)}\, ,\quad B_1 (S) = H_1 (S) - H_{-1}(S)
\end{equation}
and\footnote{See Appendix B of \cite{ABJM_QSC_Mellin} for the definition of $B_{2,3}(S)$ sums.} 
\begin{equation}
\phi_{1,1}^{\mathrm{per}} = -2 i\alpha B_1 (S)\, ,
\end{equation}
\begin{equation}
\phi_{1,0}^{\mathrm{per}} = \alpha\Big\{
\frac43 B_1(S)^2+B_2(S)+\frac{3B_3(S)+2H_3(S)-2H_{-3}(S)}{3B_1(S)}- 2 B_1 (S) (1+2\log2)\Big\}\,.
\end{equation}
The expressions for $q_2^{(0,1)}$ are too lengthy to be presented here and could be found in \cite{ABJM_QSC_Mellin}. 

Finally, let us make some comments on the generalization of the presented approach to higher loops. There we will need $\FF_{1,2}^S$ images for different products of Hurwitz functions with arbitrary indexes or fractions $\frac{1}{(u\pm i/2)^a}$ with Baxter polynomials $Q_S$. As our preliminary results show they all could be obtained algorithmically - so the current procedure seems to be systematic. The details of higher loop generalization will be the subject of one of our forthcoming publications.

\section{Anomalous dimensions}\label{AnomalousDimensions}
\noindent
Knowing the solutions of Baxter equations obtained as described in previous section the determination of the coefficients\footnote{The coefficients $m_j^{(l)}$ and $k_j^{(l)}$ are left undetermined due to QSC gauge symmetry, see \cite{ABJM_QSC_Mellin} for details.} $A_0^{(l)}$, $A_3^{(l)}$, $c_{i,k}^{(l)}$ in the ansatz for $\bP (u)$ functions introduced in section \ref{ABJM-QSC-PerturbativeSolution} goes along the same lines as in \cite{ABJM_QSC_Mellin} and we refer interested reader to that publication for further details.  This way, up to four loops we got the following results for anomalous dimensions of twist 1 operators:
\begin{equation}
\gamma (S) = \gamma^{(0)} (S) h^2 + \gamma^{(1)} (S) h^4 + \ldots
\end{equation}
where
\begin{equation}
\gamma^{(0)}(S) = 4 \left(\barH_1 + \barH_{-1} - 2\barH_i
\right)
\end{equation}
\begin{multline}
\gamma^{(1)}(S) = 16 \Big\{ 3 \barH_{-2,-1} - 2 \barH_{-2, i} - \barH_{-2,1} - \barH_{-1,-2} + 2 \barH_{-1, 2 i} - \barH_{-1,2} - 6 \barH_{i,-2} \\
+ 12 \barH_{i, 2 i} - 6 \barH_{i, 2} - 6 \barH_{2 i, -1} + 4 \barH_{2 i, i} + 2 \barH_{2 i, 1} - \barH_{1,-2} + 2 \barH_{1, 2 i} - \barH_{1,2} + 3\barH_{2,-1} \\ - 2 \barH_{2, i} - \barH_{2,1} + 2\barH_{-1, i, -1} - 2 \barH_{-1, i, 1} + 8\barH_{i, -1, -1} - 12\barH_{i, -1, i} + 4 \barH_{i, -1, 1} - 16\barH_{i, i, -1} \\
+ 16 \barH_{i, i, i} + 4 \barH_{i, 1, -1} - 4 \barH_{i, 1, i} + 2 \barH_{1,i,-1} - 2 \barH_{1,i,1} \Big\} + 8 \left(
H_{-1} - H_1
\right)\zeta_2 \\
\end{multline}
and we introduced new sums 
\begin{equation}
H_{a,b,\ldots}(S) = \sum_{k=1}^S \frac{\Re [(a/|a|)^k]}{k^{|a|}} H_{b,\ldots} (k)\, \quad H_{a,\ldots} = H_{a,\ldots} (S)\, \quad \barH_{a,\ldots} = H_{a,\ldots} (2S)\,
\end{equation}
so that for real indexes these sums reduce to ordinary harmonic sums. Imaginary indexes correspond to the generalization of the harmonic sums with the fourth root of unity factor $(\exp(i\pi/2))^n$.  The obtained  expression respects the maximal transcendentality principle \cite{N4SYM2loop4,N4SYM3loop} and is in complete agreement with previously obtained results \cite{Beccaria1,Beccaria2,ABJMQSC12loops}. This result could be actually further rewritten in terms of cyclotomic or S-sums of \cite{Ablinger3,Ablinger4} if we will extend the definition of the latter for complex values of $x_i$ parameters. It is also possible to express it in terms of twisted $\eta$-functions of \cite{cuspQSC}.

\section{Conclusion}\label{Conclusion}

In this note we have shown that it is possible to solve multiloop Baxter equations arising in quantum spectral curve problems directly in spectral parameter $u$-space by means of reducing inhomogeneous second order difference equations with complex hypergeometric functions to purely algebraic problem.
As a particular example we have considered anomalous dimensions of twist 1 operators in ABJM theory up to four loop order. The result could be expressed in terms of harmonic sums with imaginary indexes and respects the principle of maximum transcendentality. The presented technique could be further generalized both for higher loops as well as higher twists of operators under consideration.
The procedure seems to be systematic as all required $\FF_{1,2}^S$ images for different products of Hurwitz functions with arbitrary indexes or fractions $\frac{1}{(u\pm i/2)^a}$ with leading order Baxter polynomials entering at different orders of perturbative expansion in coupling constant could be obtained algorithmically. Also, similar to \cite{VolinPerturbativeSolution,ABJMQSC12loops}, where all operations close on trilinear combinations of rational, $\eta$ and $\P_k$ - functions,  all our operations close on fourlinear combinations of rational, $\eta$, $\P_k$ and $Q_S$ - functions. However, the resulted sums we get are written in a form different from generalized harmonic sums and extra work is needed to reduced them to the latter. Still, provided we already know the basis of sums entering final answer the reduction procedure could be performed by fixing coefficients in a known function basis. The advantage of the approach under development is that one can compute the required number of anomalous dimensions for different spin values in a much shorter time. All these details will be the subject of one of our subsequent publications.     

The presented techniques should be also applicable for solution of twisted $\mathcal{N}=4$ and ABJM quantum spectral curves with $\bP$ functions having twisted non-polynomial asymptotic at large spectral parameter values. The latter models are interesting in the connection with the recent progress with the so called fishnet theories  \cite{fishnet1,Isaev,fishnet2,fishnet3,fishnet4,fishnet5,fishnet6,fishnet7,fishnet8,fishnet9}. Moreover, we think that similar ideas should be also useful in the study of BFKL regime within QSC approach \cite{GromovBFKL1,GromovBFKL2,BFKLnonzeroConformalSpin} for $\mathcal{N}=4$ SYM, as the latter employs perturbative expansion in both coupling constant $g$ and parameter $w\equiv S+1$ describing proximity of Lorentz spin of operator $S$ to $-1$, so that the ratio $g^2/w$ remains fixed.

\section*{Acknowledgements}

This work was supported by RFBR grants \# 17-02-00872, \# 16-02-00943 and contract \# 02.A03.21.0003 from 27.08.2013 with Russian Ministry of Science and Education. The work of R. Lee was supported by the grant of the “Basis” foundation for theoretical physics.

\appendix

\section{Some identities for Baxter polynomials}\label{elemop}
Let us derive identities presented in the main text. We will use the
powerful approach based on the generating function for Baxter polynomials.
It has the form  \cite{ABJM_QSC_Mellin}
\begin{equation}
W\left(x,u\right)=\sum_{S=0}^{\infty}x^{S}Q_{S}\left(u\right)=(1-x)^{-\frac{1}{2}+iu}(1+x)^{-\frac{1}{2}-iu}\,.\label{eq:GF}
\end{equation}
Using this generating function, it is easy to establish various linear
relations between the Baxter polynomials. In what follows, when writing
an identity for generating function, we imply that it can be easily
checked using the explicit form \eqref{eq:GF}. 

First, expanding the identities
\[
x\partial_{x}\left(1\mp x\right)W\left(x,u\pm\tfrac{i}{2}\right)=-iux\left[W\left(x,u+\tfrac{i}{2}\right)+W\left(x,u+\tfrac{i}{2}\right)\right]
\]
in $x$, we obtain 
\begin{align}
\begin{pmatrix}Q_{S}^{\left[1\right]}\\
Q_{S}^{\left[-1\right]}
\end{pmatrix} & =\begin{pmatrix}1-i\tfrac{u}{S} & -i\tfrac{u}{S}\\
-i\tfrac{u}{S} & -1-i\tfrac{u}{S}
\end{pmatrix}\begin{pmatrix}Q_{S-1}^{\left[1\right]}\\
Q_{S-1}^{\left[-1\right]}
\end{pmatrix}\,.\label{eq:id1}
\end{align}
In particular,

\begin{equation}
Q_{S}^{\left[1\right]}-Q_{S}^{\left[-1\right]}=Q_{S-1}^{\left[1\right]}+Q_{S-1}^{\left[-1\right]}\,.\label{eq:id2}
\end{equation}
The latter identity can be used to prove

\[
Q_{S}^{\left[\pm2\right]}=Q_{S}+2\sum_{k=1}^{S}\left(\pm1\right)^{k}Q_{S-k}\,
\]
by induction (we leave this as an exercise for the reader).
Similarly, from 
\[
\left(1+x\right)\partial_{x}\left(1-x\right)\frac{W\left(x,u\right)-W\left(x,i/2\right)}{u-\frac{i}{2}}=-2iW\left(x,u\right)
\]
we obtain
\begin{align*}
\frac{\left(Q_{S}-Q_{S-1}\right)S}{u-\frac{i}{2}} & =-2iQ_{S-1}-\frac{\left(Q_{S-1}-Q_{S-2}\right)\left(S-1\right)}{u-\frac{i}{2}}\,,
\end{align*}
from which by induction we obtain
\[
\frac{Q_{S}-Q_{S-1}}{u-\frac{i}{2}}=\frac{2i}{S}\sum_{k=1}^{S}\left(-1\right)^{k}Q_{S-k}\,.
\]
This formula, in turn, can be used to prove by induction the following identity:
\begin{align*}
\frac{Q_{S}-1}{u-\frac{i}{2}} & =2i\left(-1\right)^{S}\sum_{k=1}^{S}\left(-1\right)^{k}\left(\mathrm{H}_{-1}\left(S\right)-\mathrm{H}_{-1}\left(S-k\right)\right)Q_{S-k}\,.
\end{align*}

Replacing $u\to-u$ and using the symmetry $Q_{S}\left(-u\right)=\left(-1\right)^{S}Q_{S}\left(u\right)$,
we obtain the second identity
\begin{align*}
\frac{Q_{S}-\left(-1\right)^{S}}{u+\frac{i}{2}} & =-2i\left(-1\right)^{S}\sum_{k=1}^{S}\left(\mathrm{H}_{-1}\left(S\right)-\mathrm{H}_{-1}\left(S-k\right)\right)Q_{S-k}\,.
\end{align*}

Similarly, the identities 

\[
\frac{Q_{S}^{\left[\pm1\right]}-\left(\pm1\right)^{S}}{u\pm\frac{i}{2}}=-2i\sum_{k=1}^{S}\left(\pm1\right)^{k+1}\left(\mathrm{H}_{1}\left(S\right)-\mathrm{H}_{1}\left(S-k\right)\right)Q_{S-k}
\]

can be proved.

Let us now prove some identities quadratic in the Baxter polynomials.
First, we have the following identity
\begin{gather*}
x\partial_{x}y\partial_{y}\left(x+y\right)\left(W_{x}^{\left[1\right]}W_{y}^{\left[-1\right]}+W_{x}^{\left[-1\right]}W_{y}^{\left[1\right]}\right)=-iu\left[y\partial_{y}+x\partial_{x}\right]\left(W_{x}^{\left[1\right]}-W_{x}^{\left[-1\right]}\right)\left(W_{y}^{\left[1\right]}-W_{y}^{\left[-1\right]}\right)\,,
\end{gather*}

where $W_{x}=W\left(x,u\right)$. From this identity we obtain
\begin{multline}
S_{1}S_{2}\left[Q_{S_{1}}^{\left[1\right]}Q_{S_{2}-1}^{\left[-1\right]}+Q_{S_{1}}^{\left[-1\right]}Q_{S_{2}-1}^{\left[1\right]}+Q_{S_{1}-1}^{\left[1\right]}Q_{S_{2}}^{\left[-1\right]}+Q_{S_{1}-1}^{\left[-1\right]}Q_{S_{2}}^{\left[1\right]}\right]\\
=-iu\left(S_{1}+S_{2}\right)\left(Q_{S_{1}}^{\left[1\right]}-Q_{S_{1}}^{\left[-1\right]}\right)\left(Q_{S_{2}}^{\left[1\right]}-Q_{S_{2}}^{\left[-1\right]}\right)\,.\label{eq:id3}
\end{multline}

In particular, when $S_{1}=S_{2}=S$, we have
\begin{equation}
Q_{S}^{\left[1\right]}Q_{S-1}^{\left[-1\right]}+Q_{S}^{\left[-1\right]}Q_{S-1}^{\left[1\right]}=-\frac{iu}{S}\left(Q_{S}^{\left[1\right]}-Q_{S}^{\left[-1\right]}\right)^{2}\,.\label{eq:id3-1}
\end{equation}

Let us also write the identity
\begin{equation}
Q_{S}^{\left[1\right]}Q_{S}^{\left[-1\right]}+Q_{S-1}^{\left[1\right]}Q_{S-1}^{\left[-1\right]}=\frac{-iu}{2S}\left(Q_{S}^{\left[1\right]}-Q_{S}^{\left[-1\right]}\right)\left(Q_{S}^{\left[1\right]}+Q_{S}^{\left[-1\right]}+Q_{S-1}^{\left[1\right]}-Q_{S-1}^{\left[-1\right]}\right)\label{eq:id4}
\end{equation}

which can be proved by expressing $Q_{S}^{\left[\pm1\right]}$ via
$Q_{S-1}^{\left[\pm1\right]}$ using Eq. \eqref{eq:id1}.

Next, let us prove the identity
\begin{multline}
\frac{1}{2}\left[Q_{S}^{\left[1\right]}+Q_{S}^{\left[-1\right]}+Q_{S-1}^{\left[1\right]}-Q_{S-1}^{\left[-1\right]}\right]\\
=1+\left(-1\right)^{S}-iu\left[\frac{Q_{S-1}^{\left[1\right]}+Q_{S-1}^{\left[-1\right]}}{S}+\sum_{n=1}^{S-1}\frac{1+\left(-1\right)^{n+S}}{n}\left(Q_{n-1}^{\left[1\right]}+Q_{n-1}^{\left[-1\right]}\right)\right]\label{eq:id5}
\end{multline}

Multiplying this identity by $x^{S-1}$ and summing over $S$ from
$1$ to $\infty$, we have\footnote{In the course of transformations we have to change the order of summation
	over $S$ and $n$ as follows $\sum_{S=1}^{\infty}\sum_{n=1}^{S-1}=\sum_{S=2}^{\infty}\sum_{n=1}^{S-1}=\sum_{n=1}^{\infty}\sum_{S=n+1}^{\infty}$.}
\begin{equation}\label{eq:idw}
\frac{1}{2}\left(\frac{1-x^{2}}{1+x^{2}}\right)\left[\left(1+x\right)W^{\left[1\right]}+\left(1-x\right)W^{\left[-1\right]}\right]=1-iu\int dx\left(W^{\left[1\right]}+W^{\left[-1\right]}\right)\,,
\end{equation}
where $\int dx\,\bullet$ corresponds to the operator $f\left(x\right)\to\intop_{0}^{x}d\xi\,f\left(\xi\right)$. Eq. \eqref{eq:idw} is equivalent to Eq. \eqref{eq:id5} and can be checked explicitly by evaluating both sides
of the equation and finding that they are both equal to $(1-x)^{iu}(1+x)^{-iu}$.

Now we are in position to prove the identity \eqref{eq:Qid}. Multiplying \eqref{eq:Qid} by $u Q_{S}^{\left[1\right]}Q_{S}^{\left[-1\right]}$, we arrive at
\[
Q_{S}^{\left[1\right]}Q_{S}^{\left[-1\right]}=\left(-1\right)^{S}-iu\sum_{n=0}^{S-1}\frac{1+\left(-1\right)^{n}}{2S-n}\left(Q_{S}^{\left[1\right]}Q_{S-1-n}^{\left[-1\right]}+Q_{S}^{\left[-1\right]}Q_{S-1-n}^{\left[1\right]}\right)\,.
\]
We will prove this identity by induction, with the base case easy to check. Let us prove the induction step. We have
\begin{align*}
Q_{S}^{\left[1\right]}Q_{S}^{\left[-1\right]} & =Q_{S}^{\left[1\right]}Q_{S}^{\left[-1\right]}+Q_{S-1}^{\left[1\right]}Q_{S-1}^{\left[-1\right]}-Q_{S-1}^{\left[1\right]}Q_{S-1}^{\left[-1\right]}\\
& \stackrel{\text{\eqref{eq:id4}}}{=}\frac{-iu}{2S}\left(Q_{S}^{\left[1\right]}-Q_{S}^{\left[-1\right]}\right)\left(Q_{S}^{\left[1\right]}+Q_{S}^{\left[-1\right]}+Q_{S-1}^{\left[1\right]}-Q_{S-1}^{\left[-1\right]}\right)-Q_{S-1}^{\left[1\right]}Q_{S-1}^{\left[-1\right]}\\
& \stackrel{\text{i.h.}}{=}-\frac{iu}{2S}\left(Q_{S}^{\left[1\right]}-Q_{S}^{\left[-1\right]}\right)\left(Q_{S}^{\left[1\right]}+Q_{S}^{\left[-1\right]}+Q_{S-1}^{\left[1\right]}-Q_{S-1}^{\left[-1\right]}\right)\\
&+\left(-1\right)^{S}+iu\sum_{n=0}^{S-2}\frac{1+\left(-1\right)^{n}}{2S-2-n}\left(Q_{S-1}^{\left[1\right]}Q_{S-2-n}^{\left[-1\right]}+Q_{S-1}^{\left[-1\right]}Q_{S-2-n}^{\left[1\right]}\right)\\
& =\left(-1\right)^{S}-iu\sum_{n=0}^{S-1}\frac{1+\left(-1\right)^{n}}{2S-n}\left(Q_{S}^{\left[1\right]}Q_{S-1-n}^{\left[-1\right]}+Q_{S}^{\left[-1\right]}Q_{S-1-n}^{\left[1\right]}\right)+A\,,
\end{align*}
where ``i.h.'' marks transition due to induction hypothesis, 
\begin{align*}
A & =A_{1}+A_{2}\,,\\
A_{1} & =-\frac{iu}{2S}\left(Q_{S}^{\left[1\right]}-Q_{S}^{\left[-1\right]}\right)\left(Q_{S}^{\left[1\right]}+Q_{S}^{\left[-1\right]}+Q_{S-1}^{\left[1\right]}-Q_{S-1}^{\left[-1\right]}\right)\,,\\
A_{2} & =iu\sum_{n=0}^{S-1}\frac{1+\left(-1\right)^{n}}{2S-n}\left(Q_{S}^{\left[1\right]}Q_{S-1-n}^{\left[-1\right]}+Q_{S}^{\left[-1\right]}Q_{S-1-n}^{\left[1\right]}\right)\\
&+iu\sum_{n=0}^{S-2}\frac{1+\left(-1\right)^{n}}{2S-2-n}\left(Q_{S-1}^{\left[1\right]}Q_{S-2-n}^{\left[-1\right]}+Q_{S-1}^{\left[-1\right]}Q_{S-2-n}^{\left[1\right]}\right)\,.
\end{align*}

Our goal is to prove that $A=0$. Thanks to Eqs. \eqref{eq:id5} and \eqref{eq:id2}, $A_{1}$
can be written as
\begin{multline*}
A_{1} =-\frac{iu}{S}\left(Q_{S}^{\left[1\right]}-Q_{S}^{\left[-1\right]}\right)\\\times\left(1+\left(-1\right)^{S}-iu\left[\frac{Q_{S}^{\left[1\right]}-Q_{S}^{\left[-1\right]}}{S}+\sum_{n=1}^{S-1}\frac{1+\left(-1\right)^{n+S}}{n}\left(Q_{n}^{\left[1\right]}-Q_{n}^{\left[-1\right]}\right)\right]\right)\,.
\end{multline*}

Let us now transform $A_{2}$. We shift $n\to n-2$ in the second
sum and normalize the summation limits by adding/subtracting the terms
with $n=0,1$ and $n=S$. Then we change the summation variable $n\to S-n$
and obtain
\begin{align*}
A_{2} & =iu\sum_{n=1}^{S}\frac{1+\left(-1\right)^{n+S}}{S+n}\left(Q_{S}^{\left[1\right]}Q_{n-1}^{\left[-1\right]}+Q_{S}^{\left[-1\right]}Q_{n-1}^{\left[1\right]}+Q_{S-1}^{\left[1\right]}Q_{n}^{\left[-1\right]}+Q_{S-1}^{\left[-1\right]}Q_{n}^{\left[1\right]}\right)\\
& -\frac{iu}{S}\left(Q_{S-1}^{\left[1\right]}Q_{S}^{\left[-1\right]}+Q_{S-1}^{\left[-1\right]}Q_{S}^{\left[1\right]}\right)+\frac{iu}{S}\left(1+\left(-1\right)^{S}\right)\left(Q_{S-1}^{\left[1\right]}+Q_{S-1}^{\left[-1\right]}\right)
\end{align*}
The first line can be transformed using Eq. \eqref{eq:id3} while
the second line --- using Eqs. \eqref{eq:id3-1} and \eqref{eq:id2}.
We have
\begin{align*}
A_{2} & =u^{2}\frac{1}{S}\left(Q_{S}^{\left[1\right]}-Q_{S}^{\left[-1\right]}\right)\sum_{n=1}^{S}\frac{1+\left(-1\right)^{n+S}}{n}\left(Q_{n}^{\left[1\right]}-Q_{n}^{\left[-1\right]}\right)\\
& +\frac{u^{2}}{S^{2}}\left(Q_{S}^{\left[1\right]}-Q_{S}^{\left[-1\right]}\right)^{2}+\frac{iu}{S}\left(1+\left(-1\right)^{S}\right)\left(Q_{S}^{\left[1\right]}-Q_{S}^{\left[-1\right]}\right)
\end{align*}

Comparing the final expressions for $A_{1}$ and $A_{2}$, we see,
that, indeed, $A_{1}=-A_{2}$ and thus $A=0$.

\bibliographystyle{hieeetr}
\bibliography{litr}

\end{document}